\documentclass[conference]{IEEEtran}
\IEEEoverridecommandlockouts
\usepackage{cite}
\usepackage{amsmath,amssymb,amsfonts}
\usepackage{algorithmic}
\usepackage{graphicx}
\usepackage{textcomp}
\usepackage{xcolor}
\usepackage{url}

\def\BibTeX{{\rm B\kern-.05em{\sc i\kern-.025em b}\kern-.08em
    T\kern-.1667em\lower.7ex\hbox{E}\kern-.125emX}}
\begin{document}

\title{FPGA Implementation of Sketched LiDAR for a 192 $\times$ 128 SPAD Image Sensor}


\author{%
Zhenya Zang, Mike Davies, and Istvan Gyongy\\
\textit{School of Engineering, University of Edinburgh, United Kingdom}\\
Corresponding authors: Mike Davies (Mike.Davies@ed.ac.uk), Istvan Gyongy (Istvan.Gyongy@ed.ac.uk)%
}
\maketitle

\begin{abstract}
This study presents an efficient field-programmable gate array (FPGA) implementation of a polynomial spline function–based statistical compression algorithm designed to address the critical challenge of massive data transfer bandwidth in emerging high-spatial-resolution single-photon avalanche diode (SPAD) arrays, where data rates can reach tens of gigabytes per second. In our experiments, the proposed hardware implementation achieves a compression ratio of 512× compared with conventional histogram-based outputs, with the potential for further improvement. The algorithm is first optimized in software using fixed-point (FXP) arithmetic and look-up tables (LUTs) to eliminate explicit additions, multiplications, and non-linear operations. This enables a careful balance between accuracy and hardware resource utilization. Guided by this trade-off analysis, online sketch processing elements (SPEs) are implemented on an FPGA to directly process time-stamp streams from the SPAD sensor. The implementation is validated using a customized LiDAR setup with a 192 × 128–pixel SPAD array. This work demonstrates histogram-free online depth reconstruction with high fidelity, effectively alleviating the time-stamp transfer bottleneck of SPAD arrays and offering scalability as pixel counts continue to increase for future SPADs. 
\end{abstract}


\section{Introduction}
Single-photon avalanche diode (SPAD) detectors operated with time-correlated single-photon counting (TCSPC) hardware have become prevalent in precise 3D ranging \cite{Shin2016_PhotonEfficient, LiDAR_review, li2021single} and biomedical applications \cite{SPAD_Bio_review}. Modern SPAD sensors \cite{swissSPAD, 256_256_SPAD, linospad2} are typically coupled with a field-programmable gate array (FPGA), which provides essential functions such as synchronization, readout, time-stamp decoding, and data reformatting. In time-domain applications, TCSPC hardware measures single photons' time-of-flight (ToF), where histograms accumulated from time stamps over time represent physical models containing parameters relevant to different applications. Although on-SPAD histogramming \cite{zhang201830, park202280} strategies have been reported, on-chip constructing histograms either requires substantial silicon area for storage, which becomes unsustainable as spatial resolution (pixel counts) scales up. Firmware-level time-stamp and histogram processing for emerging SPAD arrays has become more feasible with advances in semiconductor technology that allow greater hardware resources on FPGAs \cite{vornicu2017real, patanwala2019}. Existing work has explored FPGA-based direct time-stamp processing using deep neural networks (DNNs) \cite{HistogramLess_LSTM_FPGA, Li2011_VideoRate}, as well as DNN-based histogram compression \cite{Histo_Compress_Autoencoder} and compressed histogram processing \cite{Zang2023_Compact}. However, DNN-based approaches often lack generalization when experimental setups change, and pixel-wise histogram compression still consumes considerable memory resources.

In this work, we implement a time-stamp-based, histogram-free statistical compression algorithm from our earlier study~\cite{SplineSketches}, demonstrating rigorous ToF reconstruction and hardware feasibility. Here, we optimize the computation from an FPGA perspective using fixed-point (FXP) arithmetic and a design free of multiplications and nonlinear operations, thereby avoiding computational stalls that could otherwise interrupt the continuous data stream from the SPAD array.

\section{Algorithm Optimization for FPGA}
On-FPGA sketch computing for each pixel \cite{SplineSketches} in our previous paper is 
\begin{equation}
\hat{z} = \frac{1}{n} \sum_{k=1}^{n} \sum_{q=0}^{p} \phi_{i_k-q,p}(A), 
\label{eq:sketch_equation}
\end{equation}
where \( A = \left( (X \bmod T) / \Delta - i \right) \),
$\phi(\cdot)$ denotes the polynomial spline functions widely used in approximation theory \cite{unser2002splines}. Here, $n$ is the number of detected photons in the acquisition window, $T$ is the number of bins in the histogram, $\Delta = T/M$, $M$ is the size of the spline sketch, $i = 0, 1, \ldots, M-1$, and $X$ is the time stamp of each detected photon, where $X \leq T$ and $X \in [i\Delta, (i+1)\Delta)$. The parameter $p$ determines the type of polynomial spline function, either linear or non-linear. The $M$ sketch values $\hat{z}$ are then passed to a closed-form solver to generate the desired ToF bin index, which is executed off-line.  

This work focuses on computing Eq.~\ref{eq:sketch_equation} in the FPGA firmware on-the-fly, to process time-stamps directly from the SPAD array. We set $M=4$ as a trade-off between accuracy and FPGA memory constraints, which will be discussed later.  

According to our SPAD array specification \cite{192_128}, $\Delta = 4096/4 = 1024$, so
\begin{equation}
A = \left( (X \bmod 4096) / 1024 - i \right) \in [-i,\, 4 - i)
\label{eq:sketch_4096}
\end{equation}
Although non-linear and Fourier spline polynomial functions achieve higher accuracy than linear ones \cite{SplineSketches}, their hardware implementation incurs significant resource overhead and additional clock cycles, which hinder the pipelined streaming of SPAD timestamps. Therefore, we adopt a precomputed LUT-based solution for $\phi(\cdot)$, where $A$ is used as the address to fetch the corresponding values from the LUT for different spline functions within one clock cycle (CC). This avoids multiplications that would otherwise stall the timestamp processing pipeline, which operates under a one-cycle budget. The accumulation of $\hat{z}$ is implemented using BRAMs and adder logic. Since division is computationally expensive on an FPGA, photon count accumulation $n$ in Eq. \ref{eq:sketch_equation} is stored in a separate BRAM, and the final normalization is deferred to software after acquisition.  Note that although division by $1024$ could, in principle, be implemented as a right shift in hardware, our sensor's time-to-digital converters (TDCs) output is 12 bits in total (maximum value $4096$). Shifting directly by 10 bits (equivalent to division by $1024$ in \eqref{eq:sketch_4096}) would leave only 2 bits, drastically reducing accuracy and making it impossible to index a LUT with hundreds of entries. To avoid this loss of precision, instead of discretizing $A$, we retain the full integer dynamic range $[0,4096)$ and linearly map it to $2^{N_{\text{bin}}}$ entries (the LUT depth). Specifically, we define
\begin{equation}
B = \big( (X - i \cdot 1024) \bmod 4096 \big) \in [0,4096),
\label{eq:B_definition}
\end{equation}
and then map $B$ to LUT addresses using a downscale ratio $R_{dc} = 4096/N_{\text{bin}}$, such that \( B_{\text{addr}} = B / R_{dc} \), with the LUT contents generated in software for different spline polynomial functions. On FPGA, $R_{dc}$ is implemented as a binary shift by $\log_2(4096) - \log_2(N_{\text{bin}})$ bits.

\begin{figure}[t!]
    \centering
    \includegraphics[width=0.4\textwidth]{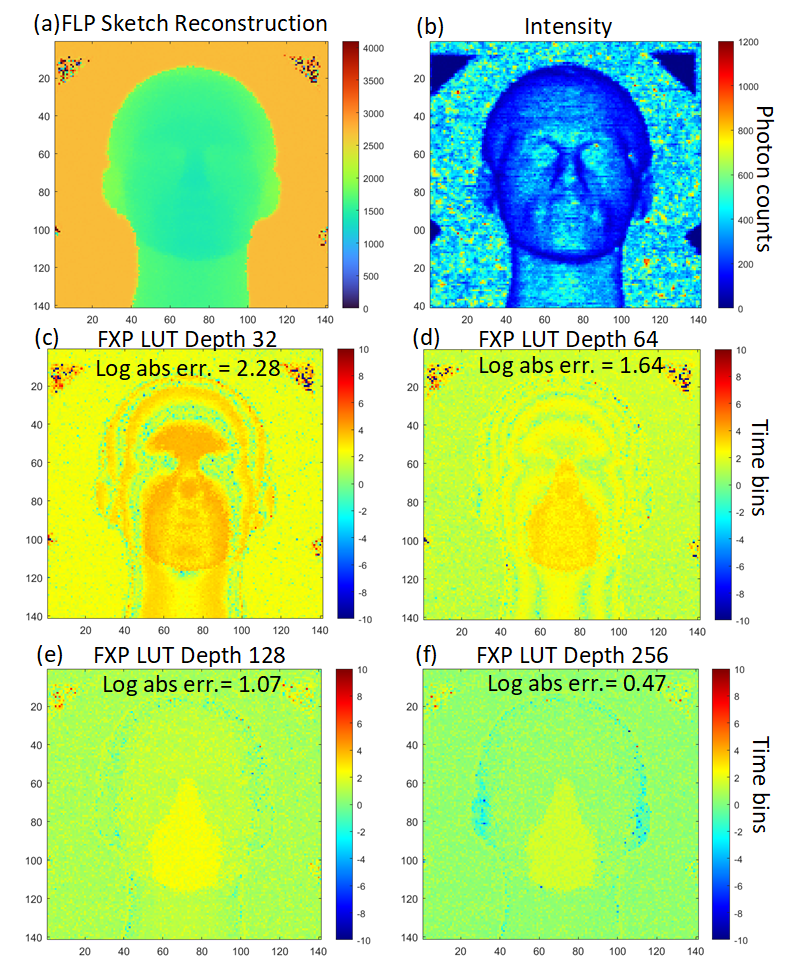}
    \caption{Simulation results. Error maps of different LUT depths with $M=4$ and $p=1$, compared with FLP reference. The dataset is taken from \cite{altmann2016lidar}.}
    \label{fig:SW_sim_results_LUT2}
\end{figure}

\section{Software Simulation and Results}
This section evaluates the effect of LUT depth and FXP bit width on reconstruction accuracy, providing guidelines for hardware implementation. We denote the FXP format as $\langle I,F \rangle$, where $I$ and $F$ are the total and fractional bit widths, respectively. We take FXP $\langle 16,7\rangle$, $M=4$, and $p=1$ as the baseline. We found that increasing $M$ to 8 or 16 produces only marginal accuracy improvements, while $M$ is recommended to be a power of two integers for hardware-friendly implementation \cite{SplineSketches}. Another important consideration is that both the on-FPGA computing logic of Eq.~\ref{eq:sketch_equation} and the FIFO storage required for $\hat{z}$ from Eq.~\ref{eq:sketch_equation} double in size when $M$ is doubled. Therefore, the above configuration is adopted for error measurements, considering the trade-off between accuracy and hardware consumption. As shown in Fig.~\ref{fig:SW_sim_results_LUT2}, using the dataset from \cite{altmann2016lidar}, which has 4613 time bins, floating point (FLP) reconstruction without the LUT approximation is taken as the reference. Four different LUT depths are evaluated. When the LUT depth reaches 256, the reconstruction achieves sub-bin accuracy with an average log error of 0.47 bins. Based on this result, we use an LUT depth of 256 when generating LUT contents and hard-coding them in ROM on the FPGA. In hardware, $M$ sketch processing elements (SPEs) are instantiated to compute Eq.~\ref{eq:B_definition}, each equipped with its own ROM for $\phi(\cdot)$ lookup. 

\section{FPGA Implementation}
\begin{figure}[t]
    \centering
    \includegraphics[width=0.4\textwidth]{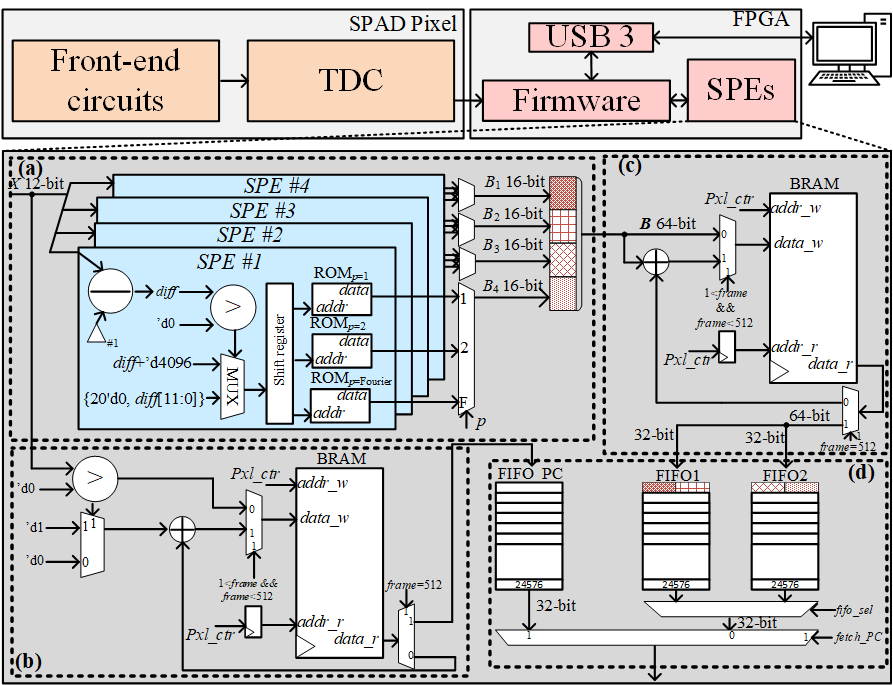}
    \caption{Hardware architecture of SPEs implementation}
    \label{fig:HW_arch}
\end{figure}

 The hardware architecture is implemented in fully parameterized Verilog. In the original firmware, the SPAD array decodes timestamps into integers per pixel for two rows ($128 \times 2$) in one batch. Time stamps from each pixel are then sequentially fed into four SPEs. The hardware architecture is shown in Fig.~\ref{fig:HW_arch}. Each pixel of the SPAD array contains front-end circuitry and a TDC that generates a stream of timestamps for the FPGA. The original firmware requires $192 \times 128$ clock cycles (CCs) to serialize timestamps from the array through a parallel-in serial-out (PISO) interface. For clarity, only one pixel is illustrated in Fig.~\ref{fig:HW_arch}. Our SPE is integrated with the firmware and completes the computation without additional pipeline bubbles or critical paths.  

The modulo processing module for Eq.~\ref{eq:B_definition} is implemented as combinational logic consisting of one adder, one subtractor, one comparator, and a shift register, as shown in Fig.~\ref{fig:HW_arch}(a). Three ROMs are instantiated to support $p=1$, $p=2$, and Fourier spline functions. After each CC, four $B$ values are computed and concatenated into a 64-bit bus. As illustrated in Fig.~\ref{fig:HW_arch}(c), one BRAM stores the accumulated $\phi_p(B)$ values for a frame, requiring $192 \times 128$ CCs to fully populate the BRAM with $192 \times 128$ entries. Since our SPAD array is single-event, detecting only one photon per frame, only two CCs are needed to write to and read from the BRAM for the accumulation $\hat{z}$ in each frame.  For frame-2, the four 16-bit $\phi_p(B)$ values fetched from the BRAM (corresponding to frame~1) are added to the current frame-2 $\phi_p(B)$ values. Once the number of frames reaches the pre-defined value of frame acquisition set by software, the accumulated 64-bit $\phi_p(B)$ results are piped out through two FIFOs as two 32-bit buses over USB~3.0. Due to the USB~3.0 specification, the FIFO entity width is fixed at 32~bits; therefore, FIFO-1 and FIFO-2 store results for $i=0,1$ and $i=2,3$, respectively. To provide the number of photon counts (PC) $n$ for Eq. \ref{eq:sketch_equation} and off-line division, another BRAM is needed to accumulated PCs of each pixel, as shown in Fig. \ref{fig:HW_arch}(b), if a time-stamp from one pixel returns with 0, indicating no photon is detected, otherwise plus one with the PCs from the previous frame. Another FIFO is instantiated for caching the PCs, as shown in Fig. \ref{fig:HW_arch}(d). Software controls \textit{fifo\_sel} and \textit{fetch\_PC} to retrieve the data through the firmware.

The maximum theoretical value of $\phi(\cdot)$ is 1. In the extreme case where a pixel generates a value of 1 in every frame, the maximum number of frames $F_{\text{max}}$ that can be accumulated is determined by the integer bit width of the FXP format. For $\langle 16,7 \rangle$, the integer width is $16-7=9$ bits, corresponding to $2^9 = 512$ frames, avoiding the overflow issue of accumulations. Thus, $F_{\text{max}}$ is configurable and represents a trade-off between FXP precision and the maximum number of frames that can be accumulated. Conventionally, our SPAD array generates $192 \times 128$ timestamps per frame. However, the SPE hardware does not output the sketch results $\hat{z}$ until the frame index reaches the predefined threshold. Consequently, the compression ratio depends on the number of accumulated frames. In our configuration, the data transfer rate is reduced by a factor of $512\times$. For frame rate, the original frame rate of the SPAD array with full temporal and spatial resolution is approximately 6,500 frames per second (fps). As discussed, the generation of the on-FPGA $\hat{z}$ does not introduce additional CC delay; however, the $\hat{z}$ frame rate is reduced by the accumulation factor on-FPGA of 512 frames. Therefore, the effective frame rate of $\hat{z}$ becomes $6{,}500/512 \approx 12.7$~ fps. To achieve a moderate video-rate depth reconstruction (e.g., 10~fps), the offline depth processing must fit within the time budget of 1/10 - 1/12.7 $\approx$ 21 ms, which establishes the lower bound for the achievable offline reconstruction rate and motivates future development of parallel hardware implementations.

\begin{table}[t]
\centering
\caption{Hardware consumption of the firmware and SPEs.}
\label{tab:hardware_consumption}
\begin{tabular}{|l|c|c|c|c|}
\hline
Components & LUT & DFF & BRAM & DSP \\ \hline
Firmware & 12,488 & 19,882 & 82 & 0 \\ \hline
SPEs     & 1,338  & 6,375  & 80 & 0 \\ \hline
\shortstack{Available\\(consumption)} &
\shortstack{134,600\\(10.27\%)} &
\shortstack{269,200\\(9.75\%)} &
\shortstack{365\\(61.13\%)} &
\shortstack{0\\(0\%)} \\
\hline
\end{tabular}
\end{table}

\begin{figure}[t]
    \centering
    \includegraphics[width=0.4\textwidth]{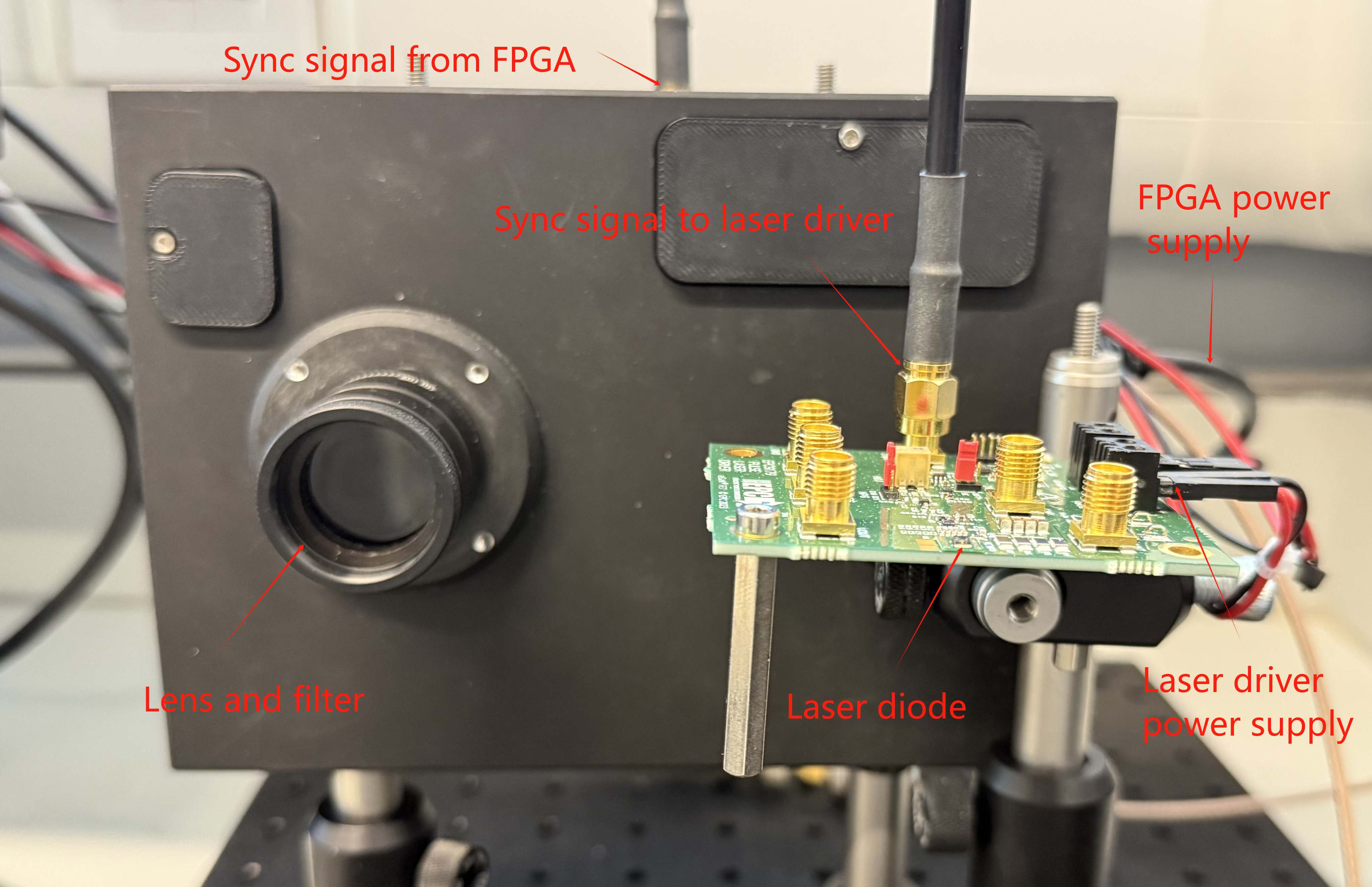}
    \caption{LiDAR setup composed of the SPAD camera and Laser.}
    \label{fig:setup}
\end{figure}

The hardware resource consumption of the original firmware and SPEs is presented in Table. \ref{tab:hardware_consumption}

\section{Real Experiment}
We use the Quanticam sensor \cite{192_128} with $192 \times 128$ pixels, configured for a temporal resolution of 40~ps. A filter and lens are mounted in front of the sensor. A pulsed laser driver (EPC9179) \cite{EPC9179} mounted with an OSRAM SPL S1L90A3 A01 surface-mounted laser diode \cite{SPLS1L90A} serves as the light source. An Opal Kelly XEM7310 FPGA \cite{XEM7310} provides a 4.54~MHz synchronous stop signal to the laser driver for pulse generation; the emission pulse width is configured to 2~ns. The exposure time of the sensor is 5 ms. The number of acquired frames is 512, as discussed.

\begin{figure}[t]
    \centering
    \includegraphics[width=0.41\textwidth]{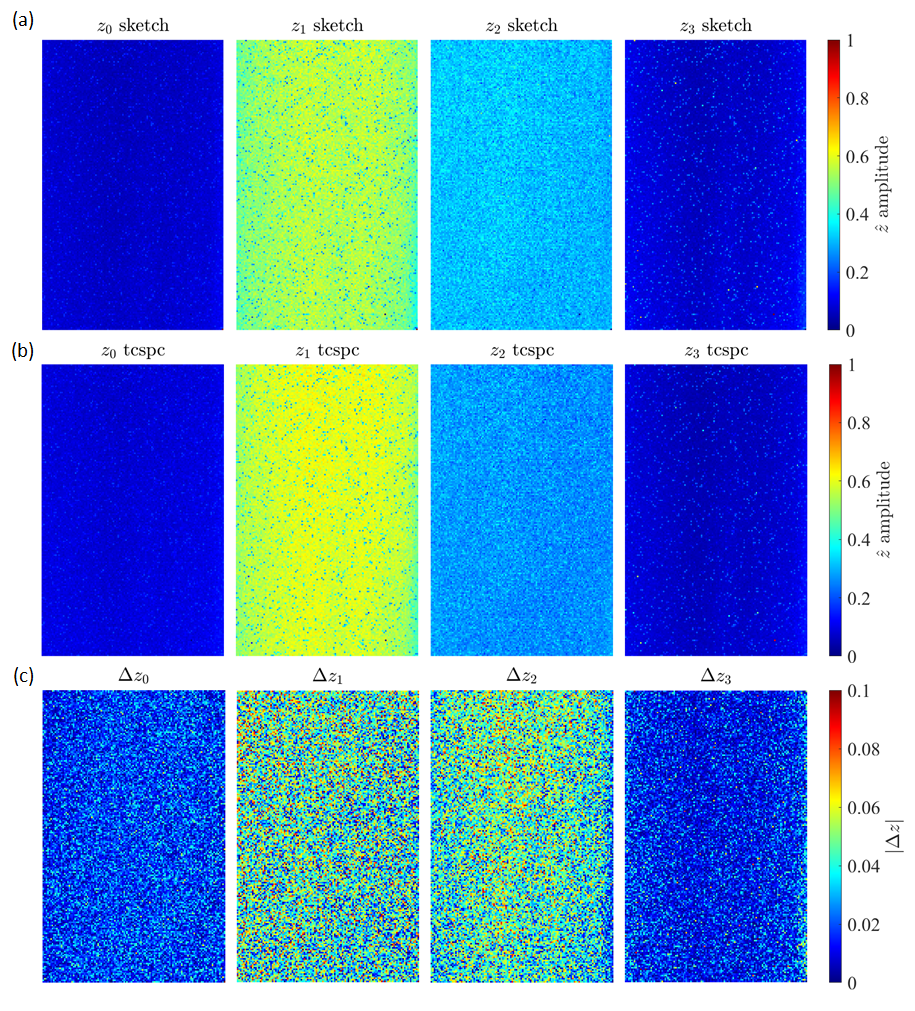}
    \caption{Optical calibration with a whiteboard placed 1 m in front of the sensor: (a) $\hat{z}$ from FPGA, (b) $\hat{z}$ from offline TCSPC histogram computation, and (c) their absolute difference.}
    \label{fig: real_board}
\end{figure}

We initially use an optical calibration board to compare the $\hat{z}$ values computed from online streams in Eq.~\ref{eq:sketch_equation} on the FPGA with those obtained from offline histogram-to-$\hat{z}$ conversion. The firmware is configured to output the values $\hat{z}$ of the SPEs and the time stamps in the TCSPC mode. As shown in Fig.~\ref{fig: real_board} (a) and (b), the online and offline $\hat{z}$ computations produce very similar $M=4$ $\hat{z}$ values, and Fig.~\ref{fig: real_board} (c) presents the array of absolute errors. 

\begin{figure}
    \centering
    \includegraphics[width=0.45\textwidth]{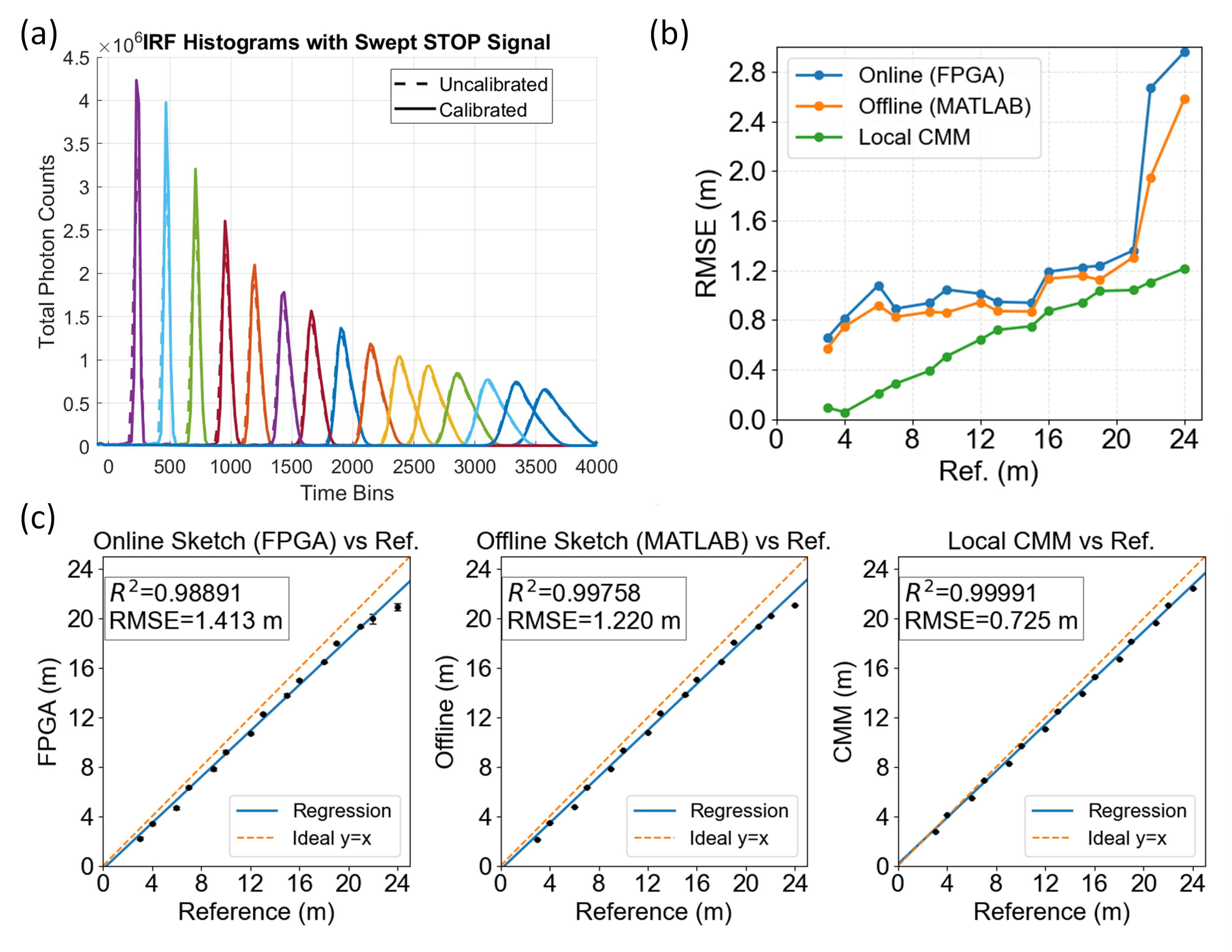}
    \caption{Accuracy analysis with varying board distances by shifting the STOP signal: (a) IRFs for 15 measurements; (b) RMSE variations from online and offline sketches; (c,d) regression results versus the local CMM reference.}
    \label{fig: distance_analysis}
\end{figure}

\begin{figure}[t]
    \centering
    \includegraphics[width=0.45\textwidth]{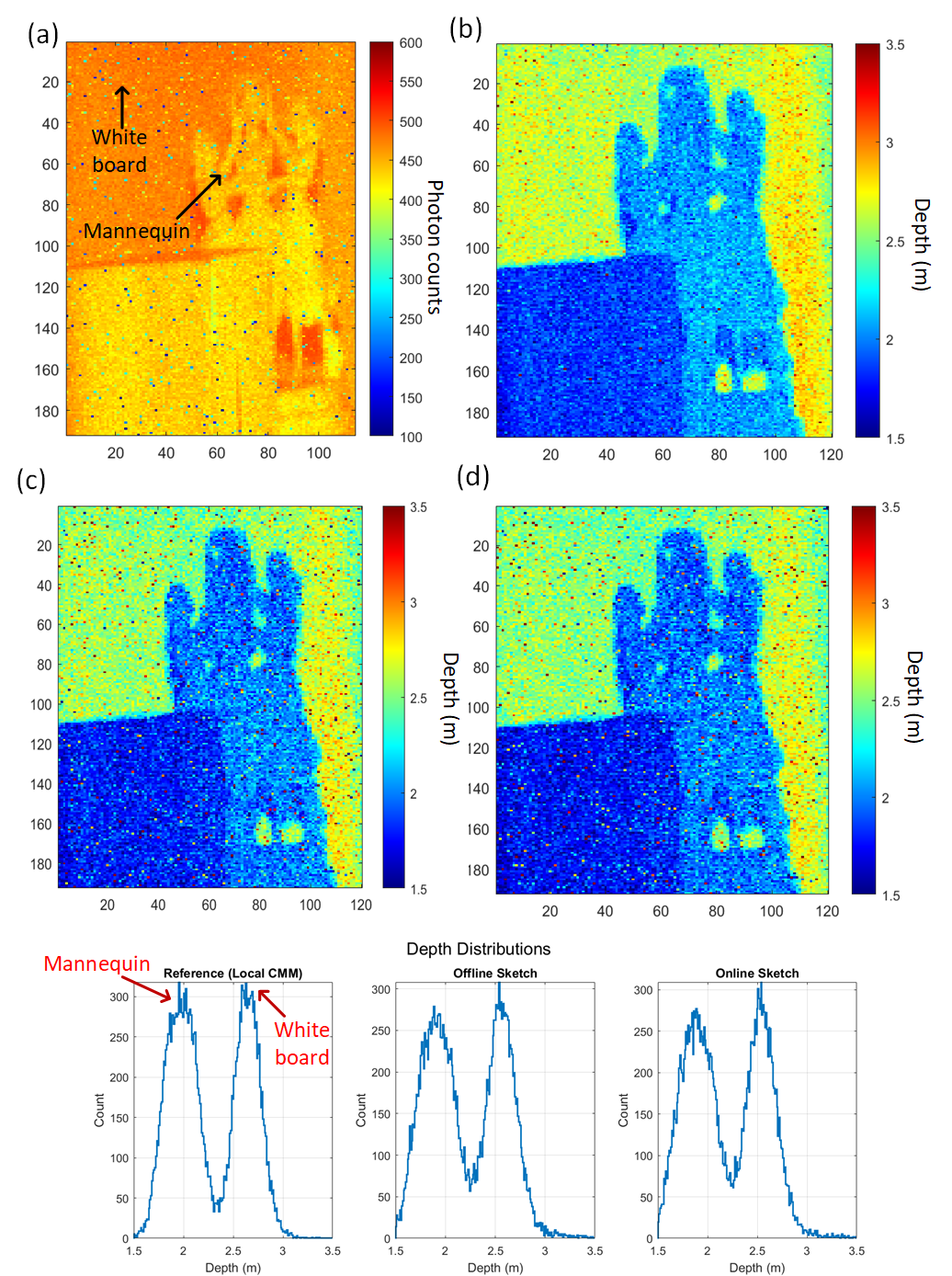}
    \caption{Real 3D-printed object sensing: (a) intensity image; (b) depth from local CMM; (c) and (d) depth reconstructed from online $\hat{z}$ (FPGA) and offline timestamp–$\hat{z}$, respectively; (e) marginal depth distributions across all pixels for the three methods.}
    \label{fig: object}
\end{figure}
Reconstruction accuracy for varying distances was investigated using the whiteboard. STOP signal was delayed in the firmware to emulate moving the whiteboard further away. The STOP signal was incrementally delayed 15 times by 10 ns each (a total of 150 ns), nearly covering the full measurement ToF range of the SPAD sensor ($\sim$163~ns). The measured histograms are shown in Fig.~\ref{fig: distance_analysis}(a), where dashed and solid curves represent uncalibrated and calibrated histograms. The pulse width increases with longer STOP delays (equivalently, larger distances, referred hereafter as “distance” for intuitive evaluation). A local center-of-mass (CMM) method within a defined temporal window was adopted as a pseudo ground truth to mitigate distortions caused by low-count time bins. Fig.~\ref{fig: distance_analysis}(b) presents the RMSE of the reconstructed depth versus distance.  RMSE trends for the FPGA-online sketch and MATLAB-offline sketch processing exhibit a similar increasing behavior. Regression analyses of the depth reconstructed from the online and offline sketch processing are shown in Figs.~\ref{fig: distance_analysis}(c). The online sketch shows a higher standard deviation than the offline version, though both have similar mean values. Both results deviate from the ideal regression line at larger distances, consistent with the RMSE trend observed in Fig.~\ref{fig: distance_analysis}(b).

We also test another scene, measuring 12 × 15 cm, which was placed 1.3 m from the setup, with a whiteboard located 1 m behind it. Fig. \ref{fig: object}(a) shows the intensity image. Fig. \ref{fig: object}(b) presents the pseudo-reference depth image from the local CMM. Figures \ref{fig: object}(c) and \ref{fig: object}(d) show the offline and online FXP depth reconstructions, respectively, using $\hat{z}$ computed in MATLAB (offline) and on an FPGA (online).
The offline result in Fig.~\ref{fig: object}(c) exhibits only marginal differences relative to the online result in Fig.~\ref{fig: object}(d). The distributions of depth values for all three methods are compared in Fig.~\ref{fig: object}(e). The peaks corresponding to the mannequin and the whiteboard are clearly distinguishable in all three distributions. Both the offline and online sketches exhibit overall closer depth values, with their entire distributions slightly left-shifted.

\section*{Conclusion}
This work presents a hardware-efficient, spline-sketch--based 3D imaging algorithm with an FPGA implementation for real-time 3D imaging, alleviating the data-transfer bottleneck between the SPAD and downstream processing hardware. The design is validated through both software simulations and real experiments. Current offline, pixel-wise postprocessing slows computation; integrating GPUs could significantly reduce runtime. Moreover, pixel-wise processing neglects spatial correlations---incorporating spatial regularization or multi-pixel models is expected to enhance reconstruction fidelity. The FPGA design currently occupies about $61.13\%$ of on-chip BRAM, constraining the maximum spline-sketch size $M$. In addition, the existing FXP format limits depth-reconstruction quality; increasing bit width would permit more frame accumulation and mitigate on-chip quantization errors. Future work will exploit the external SRAM of the Opal Kelly board to store additional $\hat{z}$ values, enabling larger $M$ and improved reconstruction accuracy.

\vspace{12pt}

\bibliographystyle{IEEEtran}
\bibliography{refs}

\end{document}